\documentclass{aa}  
\usepackage{graphicx, subfigure, amsmath}
\usepackage[breaklinks=true]{hyperref} 
\hypersetup{colorlinks=true,citecolor=blue}

\begin{document}

\title{A tunnel and a traffic jam: \\ How transition disks maintain a detectable warm dust component despite the presence of a large planet-carved gap}

\author{
P.~Pinilla\inst{1}, L.~Klarmann\inst{2}, T.~Birnstiel\inst{3}, M.~Benisty\inst{4}, C.~Dominik\inst{2}, and C.~P.~Dullemond\inst{5}
}
\institute{
Leiden Observatory, Leiden University, P.O. Box 9513, 2300RA Leiden, The Netherlands\\
\email{pinilla@strw.leidenuniv.nl}
\and
Astronomical Institute Anton Pannekoek, University of Amsterdam, PO Box 94249, 1090 GE Amsterdam, The Netherlands
\and
Harvard-Smithsonian Center for Astrophysics, 60 Garden Street, Cambridge, MA 02138, USA
\and
Univ. Grenoble Alpes, IPAG, F-38000 Grenoble, France, CNRS, IPAG, F-38000 Grenoble, France 
\and
Universit\"{a}t Heidelberg, Zentrum f\"{u}r Astronomie, Institut f\"{u}r Theoretische Astrophysik, Albert-Ueberle-Str. 2, 69120 Heidelberg, Germany
              }

 \abstract
   {Transition disks are circumstellar disks that show evidence of a dust cavity, which may be related to dynamical clearing by embedded planet(s). Most of these objects show signs of significant accretion, indicating that the inner disks are not truly empty, but that gas is still streaming through to the star. A subset of transition disks, sometimes called pre-transition disks, also shows a strong near-infrared excess, interpreted as an optically thick dusty belt located close to the dust sublimation radius within the first astronomical unit.}
   {We study the conditions for the survival and maintenance of such an inner disk in the case where a massive planet opens a gap in the disk. In this scenario, the planet filters out large dust grains that are trapped at the outer edge of the gap, while the inner regions of the disk may or may not be replenished with small grains.}
   {We combined hydrodynamical simulations of planet-disk interactions with dust evolution models that include coagulation and fragmentation of dust grains over a large range of radii and derived observational properties using radiative transfer calculations. We studied the role of the snow line in the survival of the inner disk of transition disks.}
   {Inside the snow line, the lack of ice mantles in dust particles decreases the sticking efficiency between grains. As a consequence, particles fragment at lower collision velocities than in regions beyond the snow line. This effect allows small particles to be maintained for up to a few Myrs within the first astronomical unit. These particles are closely coupled to the gas and do not drift significantly with respect to the gas. For lower mass planets (1~$M_{\rm{Jup}}$), the pre-transition appearance can be maintained even longer because dust still trickles through the gap created by the planet, moves invisibly and quickly in the form of relatively large grains through the gap, and becomes visible again as it fragments and gets slowed down inside of the snow line.}
   {The global study of dust evolution of a disk with an embedded planet, including the changes of the dust aerodynamics near the snow line, can explain the concentration of millimetre-sized particles in the outer disk and the survival of the dust in the inner disk if a large dust trap is present in the outer disk. This behaviour solves the conundrum of the combination of both near-infrared excess and ring-like millimetre emission observed in several transition disks.}


\titlerunning{A tunnel and a traffic jam}
\authorrunning{P.~Pinilla et al.}
\maketitle

\section{Introduction}     \label{introduction}
In recent decades, the observations of transition disks has lead to a major step forward in our understanding of the physical processes of circumstellar disk evolution and clearing \citep[e.g.][]{najita2007, williams2011, espaillat2014}. Transition disks were originally characterised by their spectral energy distributions (SEDs), which show weak near- and mid-infrared excess emissions \citep{strom1989, skrutskie1990} but substantial excess beyond $\gtrsim 20\mu$m. This type of SED  indicated the presence of a dust-depleted cavity in the inner regions of the disk, which was recently confirmed by spatially resolved images in the submillimetre continuum of the largest cavities \citep[e.g.][]{andrews2011, vandishoeck2015}. Observations with \textsl{Spitzer} showed a broad diversity in  the  morphology of the SEDs of transition disks \citep[e.g.][]{cieza2007}. For instance, a subset of disks shows a strong excess in the near-infrared  (NIR) wavelength range \citep[e.g. LkCa~15 or UX~Tau~A,][]{espaillat2010} similar to the median SED of disks in the Taurus star-forming region \citep{alessio1999}. A compact, optically thick inner disk close to the sublimation radius where dust is destroyed (at temperatures higher than ${\sim} 1500$K) has been evoked to reproduce this NIR excess \citep{espaillat2010}. In addition, interferometric observations at the NIR have spatially resolved a compact inner disk in some transition disks \citep[e.g.][]{benisty2010, olofsson2013}. These disks are also known as  pre-transition disks, because they are thought to be at an earlier evolutionary stage than transition disks, whose cavities are completely empty of dust. Demographics of the well-known transition disks have revealed that about half of the transition disk population have a NIR excess \citep[see Table 1 in][]{espaillat2014}. 

Gas inside the dust cavities of transition disks has been detected \citep[e.g.][]{bergin2004, salyk2007, pontoppidan2008}, although it also appears depleted, albeit by a lower factor and within a smaller cavity than the millimetre dust \citep{bruderer2014, zhang2014, vandermarel2015}. The accretion rates in transition disks have values similar to those measured in full disks \citep[${\sim}10^{-10}-10^{-7}\rm{M}_{\odot}\rm{yr}^{-1}$, e.g., ][]{najita2007, najita2015, cieza2008, sicilia2010}, which means that significant quantities of gas are still present in the inner regions. 

The physical mechanism(s) responsible for the observed dust and gas properties of transition disks is debated and it is still uncertain whether multiple processes simultaneously occur or whether a single process dominates the disk evolution. To date, models have mostly focused on the properties of the outer disk and it is still unclear why some disks have an optically thick inner disk, and under which conditions it is maintained. For instance, \cite{birnstiel2012b} demonstrated that grain growth alone can reproduce the near-/mid-IR dips of the SEDs of transition disks, but fails to reproduce the observed cavities in the submillimetre continuum. Photoevaporation from far-UV and X-ray radiation can dissolve the inner disk on short time scales (${\sim}$0.5 Myr) once accretion stops \citep[e.g.][]{alexander2006}, and can explain transition disks with small cavities and low accretion rates \citep[e.g.][]{najita2007, cieza2008, owen2011, owen2012}. However, since accretion effectively  stops in photoevaporating disks, such models cannot explain transition disks with signatures of on-going accretion or the large fractions of large cavity sizes.  Also, any dust signatures linked to the inner disk should then disappear on an accretion time scale. An alternative explanation invokes dead zones. These are regions of low ionisation fraction that hinder the magneto-rotational instability (MRI) needed to drive the turbulence necessary for the transport of angular momentum \citep{balbus1991}. Because the accretion rate onto the star depends on disk turbulence, at the edge of a dead-zone a pile-up of gas material forms, which can also lead to gaps as observed at millimetre wavelengths \citep{regaly2012, flock2015}. 

Dynamical clearing by planet(s) or massive companions can also be a possibility for the origin of the cavities  \citep[e.g.][]{goldreich1980, crida2006, varniere2006}. Dust dynamics in the presence of embedded planets can lead to dust filtration \citep[e.g.][]{paardekooper2006, rice2006}, creating a spatial segregation of small and large grains. This segregation depends on the planet mass and disk turbulence, which influence the gap shape and particle diffusion through the gap \citep[e.g.][]{zhu2012, pinilla2012}. The combination of grain growth/fragmentation and filtration can predict two major observational features: a ring-like emission at mm wavelengths (tracing large grains), and either a smaller cavity, or no cavity at all in NIR scattered light images (tracing small grains) \citep{dong2012, dejuanovelar2013, follette2013, garufi2013}. If the embedded planets are not too massive, small grains are not trapped in the pressure bump at the outer edge of the gap, and can move inwards through the gap. In the inner disk, these particles can grow again and be lost towards the star as a result of their rapid inward drift. In mm and sub-mm images, this component is not prominent, because it is only a fraction of the dust material that makes it past the pressure bump, and because the fast drift leads to lower surface densities according to $\dot{M}=2\pi\Sigma_d r v_{\rm{drift}}$, where $\Sigma_d$ is the dust surface density and $v_{\rm{drift}}$ the radial drift velocity. In addition, if grain growth is very efficient, the grains might even become too large for efficient mm wavelength radiation emission.  To date, there is no conclusive answer for the reappearance of the inner disk.

In this paper, we  study the conditions for the reappearance of such an inner disk after a massive planet has opened a gap in the disk. We investigate the role of the water snow line for the survival of the inner disk in transition disks when a massive planet is embedded in the outer disk regions (20~AU) and explore what conditions are needed to produce both a significant NIR excess and a ring-like feature in the submillimetre. The snow line affects the dynamics of the dust particles because grains with ice mantles (located outside of the snow line) are expected to stick more efficiently than pure silicates \citep{chokshi1993, schafer2007, wada2009, gundlach2015}, and therefore the velocity thresholds \citep[e.g][]{dominik1997} for destructive collisions change (fragmentation velocities $v_{\rm{frag}}$) at the snow line location. \cite{birnstiel2010} showed the effect of changing $v_{\rm{frag}}$ from $1~\rm{m}~\rm{s}^{-1}$ inside versus $10~\rm{m}~\rm{s}^{-1}$ outside the snow line. Beyond the snow line,  grains grow as a result of coagulation (from ${\sim}$cm sizes near to the snow line to ${\sim}$mm-sized grains further out), they start to decouple from the gas, and therefore they drift efficiently inwards owing to their friction with the gas \citep[e.g][]{weidenschilling1977}. However, inside the snow line, because of the higher fragmentation velocities, the particles remain sufficiently small and so are not influenced by the radial drift, and they remain closely coupled to the gas, producing an accumulation of dust near the snow line\footnote{For the purpose of this study, it is irrelevant whether aggregates remain intact while losing their ice component after passing through the snow line or whether the ice sublimation disintegrates the aggregates into silicate monomers that then grow again from scratch.}. In the case of planets embedded in the disk, the amount of dust at the snow line location would depend on how much dust is filtered by the planet in the outer regions. 

\begin{figure}
 \centering
   \includegraphics[width=9.0cm]{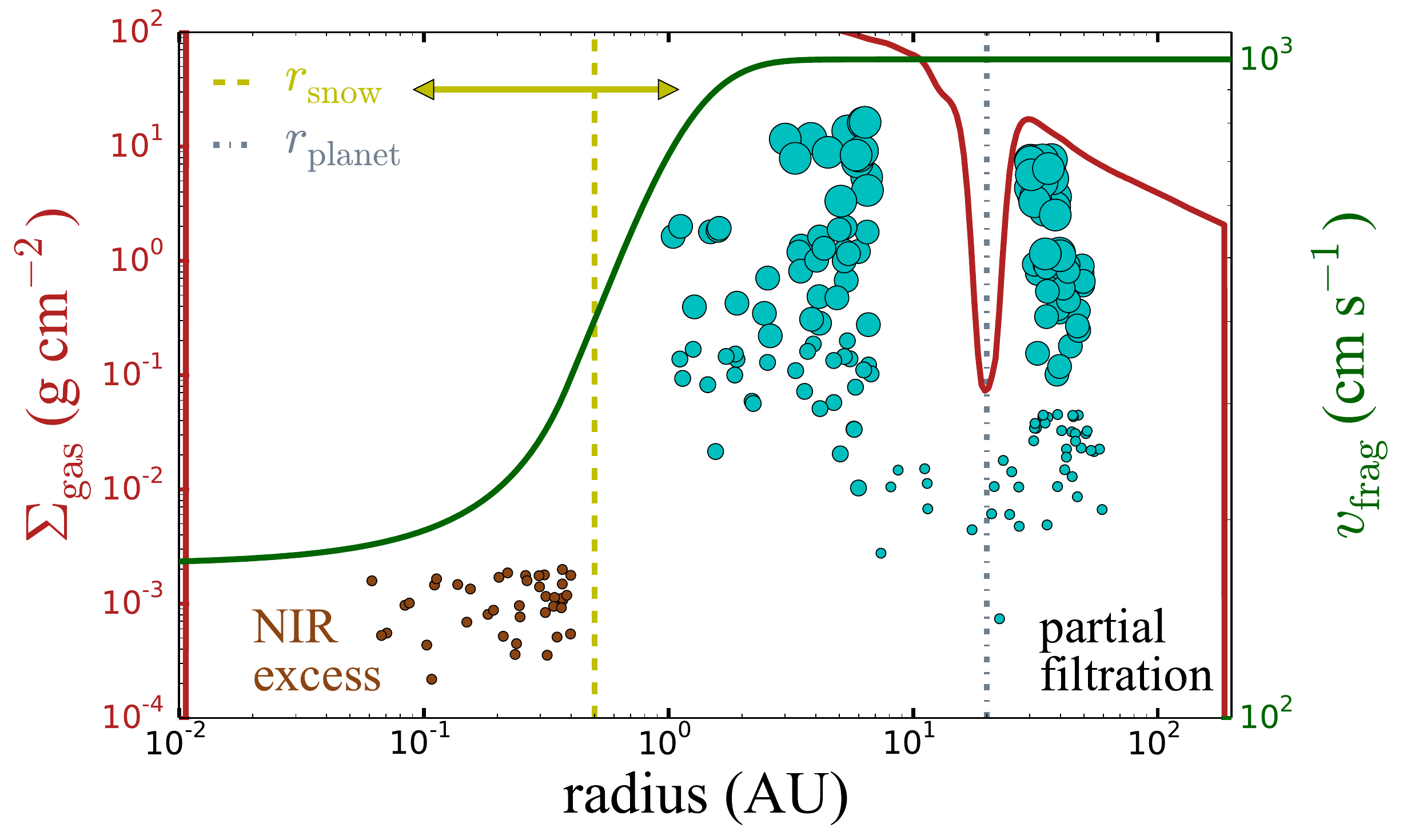}
   \caption{Sketch of particle trapping triggered by a planet located in the outer disk (20AU, vertical dash-dotted line), and the effect of the snow line on the dust distribution. The partial filtration of dust particles at the outer edge of the planetary gap and the change of the fragmentation velocity of dust particles near the snow line ($v_{\rm{frag}}$ from $1~\rm{m}~\rm{s}^{-1}$ inside versus $10~\rm{m}~\rm{s}^{-1}$ outside) allows the constant replenishment of small particles in the inner disk. The vertical dashed line represents the snow line position, which can vary within a certain range (horizontal arrow for a typical T Tauri star).}
   \label{cartoon}
\end{figure}

In this work we combine the effect of a massive planet embedded in the disk together with the change of dust dynamics near the snow line.  Figure~\ref{cartoon} represents the main idea addressed in this paper: particle trapping triggered by a planet located in the outer disk, together with the impact that the water snow line has on the dust distribution. In Sect.~\ref{method} we explain the models used. In Sects~\ref{results} and \ref{discussion} we present the main results and discussion. We finish with the main conclusions in Sect.~\ref{conclusion}.

\section{Method}     \label{method}
In this work we combined hydrodynamical simulations of planet-disk interaction, dust evolution, and radiative transfer models as in \cite{pinilla2015}. 

For the hydrodynamical models, we used the 2D version of the fast advection hydrodynamical code \texttt{FARGO} \citep{masset2000}, with the same set-up as in \cite{pinilla2012}, i.e. a non-migrating planet interacting with a flared disk, with an initial gas density profile of $\Sigma_g\propto r^{-1}$, and a disk scale height that varies with radius as $h/r \propto r ^{1/4}$. In terms of the planet position ($r_{\rm{planet}}$), the radial grid logarithmically covers $[0.1-7.0]\times r_{\rm{planet}}$ with 512 cells, and 1024 linearly spaced cells for the azimuthal grid (from 0 to $2\pi$). We performed the simulations until 1000 planetary orbits assuming three values for the planet-to-star mass ratio of $q=[1.0\times10^{-3}, 5.0\times10^{-3}, 1.5\times 10^{-2}]$, which corresponds to 1, 5 and 15~$M_{\rm{Jup}}$ mass planets around a solar-type star. The disk is assumed to have a mass of $M_{\rm{disk}}=55~M_{\rm{Jup}}$. We are mainly interested in the radial distribution of grains and want to apply dust evolution including growth, fragmentation, and erosion of particles as introduced by \cite{birnstiel2010}. This can only be modelled in the radial direction. Therefore we azimuthally averaged the gas surface density of the last 100 orbits and assumed this profile as background gas density for the dust evolution. For this work, we assumed a single value of the disk viscosity, parametrised by $\alpha_{\rm{turb}}$ \citep{shakura1973} and taken to be $\alpha_{\rm{turb}}=10^{-3}$.

For the dust evolution, we used the models explained in \cite{birnstiel2010}, assuming a typical T Tauri star. The planet is located at 20AU, and therefore the radial grid from the hydrodynamical simulation only covers a region from 2~AU to 140~AU. Since we are also interested in the inner disk, we extrapolated the results from \texttt{FARGO} to a grid from 0.1 AU (close to the dust sublimation radius) to 200~AU, where the gas surface density profile follows the initial condition. In these models, radial drift, turbulent mixing, and gas drag are taken into account for the grain growth and dynamics. Initially all particles are 1~$\mu$m in size and distributed with a dust-to-gas mass ratio of 1/100. The grid of particle sizes ranges from 1~$\mu$m to 200~cm, divided into 180 logarithmic size bins.  

To quantify the coupling of the dust particles to the gas, we refer to the Stokes number ($\rm{St}$), which is defined as a dimensionless stopping time $\rm{St}=t_s\times\Omega$ (where $\Omega$ is the orbital period, $t_s$ is the stopping time of a particle), and in the Epstein regime (i.e. $\lambda_{\rm{mfp}} \geq 4/9 a$, $a$ being the particle size and $\lambda_{\rm{mfp}}$  the mean free path of the gas molecules),  $\rm{St}=$ $a \rho_s \pi /2\Sigma_g$ ($\rho_s$ being the volume density of a grain size and taken to be 1.2~$\rm{g~cm}^{-3}$). 

The fragmentation threshold velocities for ices is about ten times higher than for silicates \citep[e.g.][]{blum2008}. By performing numerical simulations, \cite{wada2013} showed that $v_{\rm{frag}}$ can be $10-80~\rm{m}~\rm{s}^{-1}$ for ice aggregates and $1-10~\rm{m}~\rm{s}^{-1}$ for silicates. Experimental work on collisions of spherical, micron-sized water-ice particles showed lower values:  ${\sim}1-2~\rm{m}~\rm{s}^{-1}$ for silicates and ${\sim}10-30~\rm{m}~\rm{s}^{-1}$ for ice monomers of $\gtrsim 1\mu$m size \citep{gundlach2015}. In this work we assume values of $v_{\rm{frag}}$ from $1~\rm{m}~\rm{s}^{-1}$ inside to $10~\rm{m}~\rm{s}^{-1}$ outside the snow line. We assume this change of $v_{\rm{frag}}$ to occur smoothly within a range of 2 ~AU. This transition is chosen to avoid numerical problems in the simulations and to mimic a smooth vertical distribution of the temperature and distribution of ices in a turbulent disk \citep{furuya2014}.   \cite{min2011} showed that the location of the snow line ($r_{\rm{snow}}$) strongly depends on the dust opacities and the disk accretion rate, and found that for a typical T Tauri star and accretion rates of ${\sim}10^{-10}-10^{-7}M_{\odot}\rm{yr}^{-1}$,  $r_{\rm{snow}}$ in the mid-plane is ${\sim}0.5 - 2$~AU. \cite{mulders2015} included large grains in the calculation of the snow line location, finding that it can be two times closer to the star. In our models, the location of the snow line depends on the dust midplane temperature profile, and it is assumed that $r_{\rm{snow}}$ is at $T_{\rm{snow}}{\simeq}200$~K \citep[e.g.][]{lecar2006}. We have assumed two different temperature profiles, such that $r_{\rm{snow}}$ is at 0.4~AU (reference model) or at 1.2~AU. In summary, the reference models consider three different planet masses (1, 5, and 15~$M_{\rm{Jup}}$ mass planets around a $1~M_{\odot}$ star). For the 1 and 5 $M_{\rm{Jup}}$ cases, an additional model is shown where the snow line is further out.

In order to compare with observations, we input the vertically integrated dust density distributions from the models after several million years  ($[1,5]$~Myr) to Monte Carlo radiative transfer code MCMax  \citep{min2009}. For our models, MCMax calculates the density and temperature structure in a 2D setup $(r, z)$, but does 3D radiative transfer following the description in \cite{bjorkman2001}.  The $\alpha_{\rm{turb}}$ parameter is the same in all three parts of the computation: hydrodynamics (viscosity), dust evolution, and radiative transfer (diffusion and collision velocities, and vertical distribution). As scattered light contributes significantly to the NIR flux, we also include anisotropic scattering \citep{min2012, mulders2013}. For the dust composition, we adopted the values from \cite{ricci2010}, i.e. porous spheres made of astronomical silicates (${\sim}$10\% in volume), carbonaceous materials (${\sim}$20\%), and water ice (${\sim}$30\%). Dust porosity can vary during the dust evolution process, for example by porous coagulation \citep{ormel2007a, okuzumi2009, zsom2010}, pressure compaction \citep{kataoka2013}, or erosion of the small grains \citep{krijt2015}.  However, our interest is focused on the inner parts of the disk, where particles already have passed the regions around a planet where shocks and/or heating by the planet might cause further compaction of the grains. Furthermore, in the inner few AUs, sintering might affect the structure of the grains \citep{sirono2011}. In this study, we assume the particle porosity to be constant. Allowing for a varying porosity would cause particles to reach larger Stokes numbers \citep[e.g.][]{krijt2015}, which would amplify the effects discussed in this paper. Hence, assuming a constant porosity is a conservative approximation in this scenario.

\begin{figure*}
 \centering
   	\includegraphics[width=16.0cm]{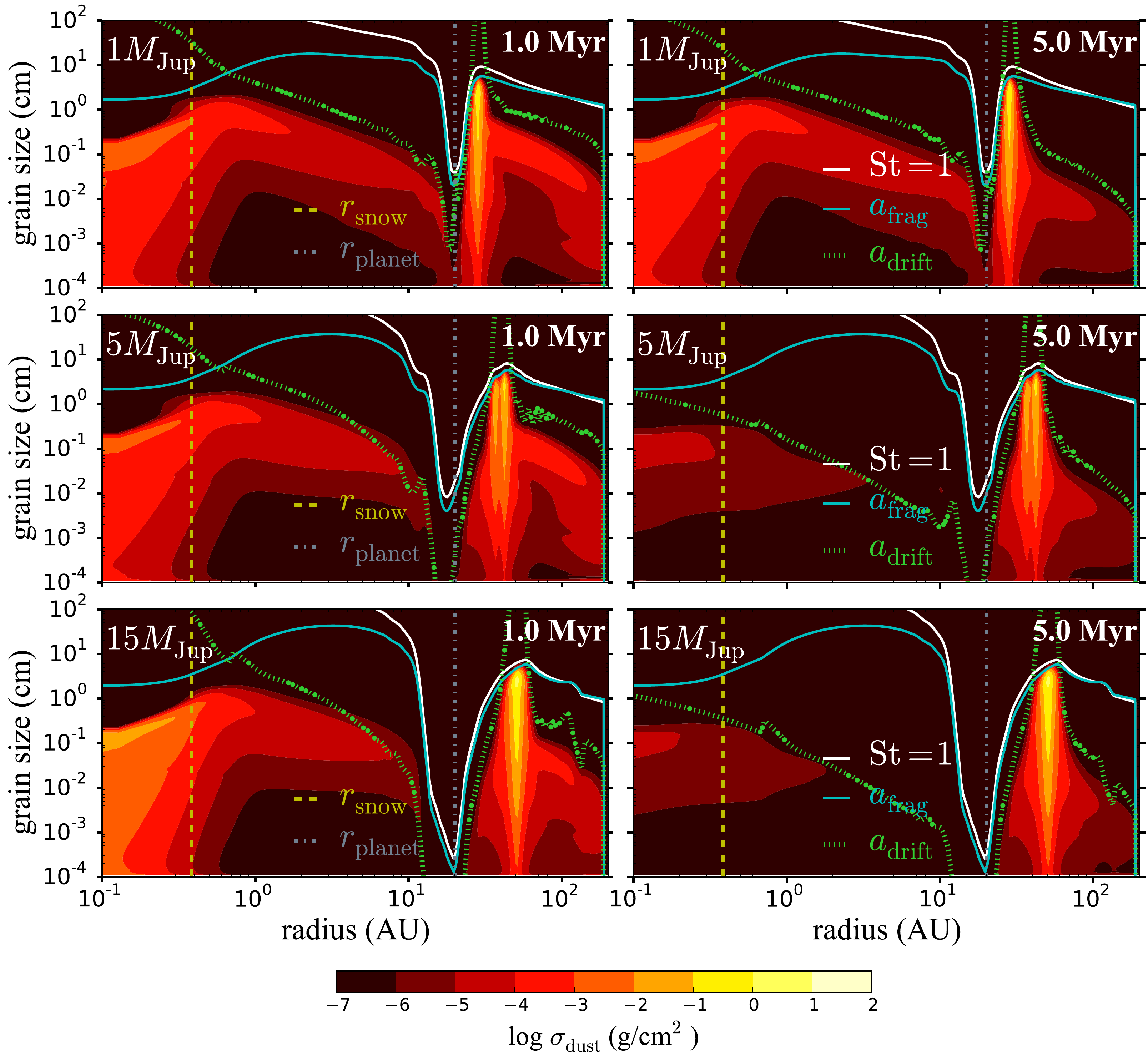}
   \caption{Vertically integrated dust density distribution after 1 (left) and 5 (right) Myrs of evolution, when a massive planet ($1, 5,~\rm{and}~15~M_{\rm{Jup}}$, from top to bottom)  is embedded in the disk at 20~AU (vertical dash-dotted line). At the location of the snow line (vertical dashed line at ${\sim}0.4$~AU), the fragmentation velocity changes from $1~\rm{m}~\rm{s}^{-1}$ to $10~\rm{m}~\rm{s}^{-1}$ within a range of ${\sim}2$~AU. The solid white line corresponds to $\rm{St}=1$, which is proportional to $\Sigma_g$. The solid blue line corresponds to the maximum grain size due to fragmentation (Eq.~\ref{afrag}), while the dotted green line represents the drift barrier (Eq.~\ref{adrift}).}
   \label{dust_distribution}
\end{figure*}

\section{Results}     \label{results}

In this section, we present the results of particle trapping triggered by a planet (1, 5, and 15~$M_{\rm{Jup}}$) located at 20~AU, together with the impact that the water snow line has on the fragmentation velocity of particles, thus on the final dust density distributions. For the explanation of the results, we first introduce the definitions of the maximum grain size due to fragmentation or radial drift as explained in \cite{birnstiel2012a}.

\subsection{Maximum grain size due to fragmentation and radial drift}

In dust evolution models, when particles start to decouple from the gas the dominant sources for their relative velocities are turbulence and radial drift, both depend on the Stokes number. The maximum velocities  by turbulence are given by \cite{ormel2007b},

\begin{equation}
v_{\rm{turb}}^2=\frac{3}{2}\alpha_{\rm{turb}}\rm{St}~c_s^2
\label{v_turb}
\end{equation}

\noindent with $c_s$ being the sound speed;  the drift velocities are given by \citep{weidenschilling1977}

\begin{equation}
v_{\rm{drift}}=\frac{1}{\rm{St}^{-1}+\rm{St}}\frac{\partial_r P}{\rho_g \Omega},
\label{v_drift}
\end{equation}

\noindent with $P$ the disk gas pressure, $\rho_g$ the volume gas density, and $\Omega$  the Keplerian frequency. When dust fragmentation mainly occurs because of turbulence, the maximum grain size in the Epstein regime (fragmentation barrier, $a_{\rm{frag}})$ is given by 

\begin{equation}
	a_{\rm{frag}}=\frac{2}{3\pi}\frac{\Sigma}{\rho_s \alpha_{\rm{turb}}}\frac{v_{\rm{frag}}^2}{c_s^2}.
  \label{afrag}
\end{equation}

However, if particles reach values of  $\rm{St}$ close to unity prior to the fragmentation barrier (Eq.~\ref{afrag}), they can only reach certain sizes ($a_{\rm{drift}}$) before they drift (drift barrier), which in the Epstein regime is given by

\begin{equation}
	a_{\rm{drift}}=\frac{2 \Sigma_d}{\pi\rho_s}\frac{v_K^2}{c_s^2}\left \vert \frac{d \ln P}{d\ln r} \right \vert^{-1}.
  \label{adrift}
\end{equation}

\subsection{Impact of the water snow line on the dust distribution}

Figure~\ref{dust_distribution} illustrates the vertically integrated dust density distribution after 1 and 5~Myr of evolution, when a massive planet ($1, 5,~\rm{and}~15~M_{\rm{Jup}}$)  is embedded in the disk at 20~AU. At the location of the water snow line, the fragmentation velocity ($v_{\rm{frag}}$) changes from $1~\rm{m}~\rm{s}^{-1}$ to $10~\rm{m}~\rm{s}^{-1}$ within a range of ${\sim}2$~AU. This transition is due to the lack of ice mantles in dust grains inside the snow line, which decreases the sticking efficiency. Figure~\ref{dust_distribution} also shows the particle sizes that correspond to $\rm{St=1}$, the  maximum grains size due to fragmentation ($a_{\rm{frag}}$) or radial drift ($a_{\rm{drift}}$). For simplicity Eqs.~\ref{afrag} and \ref{adrift} assumed the Epstein drag regime; however in the simulations the first Stokes drag regime is also considered.

At the outer edge of the planetary gap, where the gas surface density increases outwards and the pressure gradient is hence positive, the particles drift outwards and they are trapped at those locations (Eq.~\ref{v_drift}). Irrespective of the planet mass, the drift barrier at the outer edge of the gap is irrelevant because the drift velocities are reduced or completely suppressed.  In the pressure bump, the maximum grain size is determined by the fragmentation barrier (Eq.~\ref{afrag}) and particles can reach centimetre sizes, but small particles are continuously reproduced. For a more massive planet, the location of the pressure maximum (and therefore the peak of mm dust concentration) occurs further away at 30, 44, and 54~AU for $1, 5,~\rm{and}~15~M_{\rm{Jup}}$, respectively, as previously shown by \cite{pinilla2012}. The radial range in which mm-sized grains are concentrated becomes wider. For the $5~M_{\rm{Jup}}$ case, the second peak is caused by a transient density feature that in reality would be short-lived and would not function as a pressure maximum that can effectively concentrate particles.  This issue has no impact on our results since we focus on the inner disk.

In the regions inside the planetary gap ($r\lesssim10$~AU) where the gas surface density steeply decreases outwards (and therefore the pressure gradient is negative), the maximum grain size is determined by radial drift (Fig.~\ref{dust_distribution}), and only close to the snow line ($r{\sim} 0.7-0.8$~AU) does the fragmentation barrier drop below the drift barrier due to the change in $v_{\rm{frag}}$ at $r_{\rm{snow}}$. With the decrease in $v_{\rm{frag}}$ at $r_{\rm{snow}}$, the maximum grain size is a few centimetres and the Stokes number of the particles at that location is much lower than unity  ($\rm{St}{\sim}1\times10^{-3}-5\times10^{-3}$), thus  they are better coupled to the gas, their radial drift velocities are lower (Eq.~\ref{v_drift}), and they move with at a similar speed to the gas \citep[see also][]{brauer2008}. If $v_{\rm{frag}}$ does not change at $r_{\rm{snow}}$, the maximum grain size is dominated by radial drift, reaching values of the Stokes number of around $0.1$, which would lead to very high drift velocities  (${\sim}1000~\rm{cm~s}^{-1}$) depleting the inner disk on very short time scales (${\sim}$100-1000 yr). Therefore, when the inner disk is constantly resupplied from the outer disk, the changes of  $v_{\rm{frag}}$ at  $r_{\rm{snow}}$ allows replenishment of warm, small, and optically thick dust near $r_{\rm{snow}}$, for which the drift velocities are slow. 

In the case of a $1~M_{\rm{Jup}}$ mass planet, the micron-sized particles (1-10~$\mu$m), are not perfectly trapped at the outer edge of the planet-gap, and they move continuously through the gap via turbulent diffusion.  This partial filtration of dust grains supports a constant replenishment of dust from the outer to the inner disk, which together with variations of  $v_{\rm{frag}}$ at  $r_{\rm{snow}}$ extends the lifetime of a detectable inner disk reaching the life of the disk (5~Myr). 

In the cases of $5~M_{\rm{Jup}}$ and $15~M_{\rm{Jup}}$ mass planets, there is more efficient trapping of particles at the outer edge of the planetary gap, filtering all particle sizes (from 1$\mu$m, which is the minimum grain size in our setup). The dust  inside the planetary gap ($r\lesssim10$~AU) is the dust that was initially within the gap. These particles grow, drift and fragment when they reach the snow line. Although there is a considerable amount of dust in the inner disk for all cases ($1, 5,~\rm{and}~15~M_{\rm{Jup}}$) after 1~Myr,  after 5~Myr of evolution the  inner disk is practically depleted of dust for a disk hosting a $5~M_{\rm{Jup}}$ and a $15~M_{\rm{Jup}}$ mass planet. The dust density distribution after 5~Myr of evolution for $r\lesssim10$~AU is almost identical for $5~M_{\rm{Jup}}$ and $15~M_{\rm{Jup}}$ because the effectiveness of trapping at the outer edge of the planet gap is similar \citep{pinilla2012}. However, the peak of the concentration of mm-sized grains  in the outer disk differs by around ~10 AU (44~AU for a $5~M_{\rm{Jup}}$ versus 54~AU for a $5~M_{\rm{Jup}}$ mass planet). Thus, the only effect of growing the planet mass beyond $5~M_{\rm{Jup}}$ is to increase the separation between the planetary orbit and the emission ring seen at mm and sub-mm wavelengths.

\begin{figure}
 \centering
   	\includegraphics[width=8.5cm]{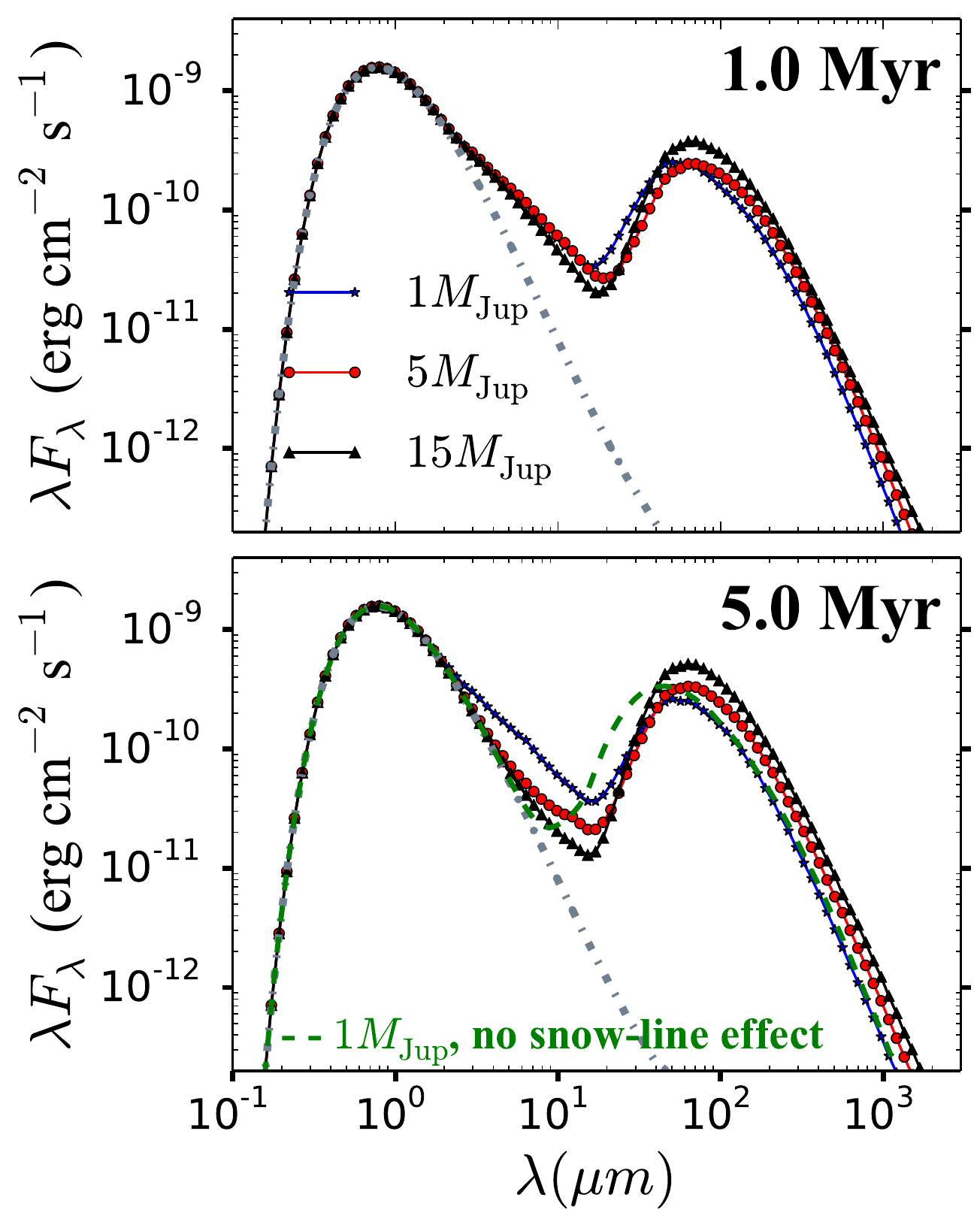}
   \caption{SEDs obtained from the dust density distribution after 1~Myr (upper panel) and 5~Myr (lower panel) for the case of a $1, 5,~\rm{or}~15~M_{\rm{Jup}}$ mass planet embedded at 20~AU. In the lower panel the dashed line corresponds to the SEDs obtained by assuming a planet mass of $1~M_{\rm{Jup}}$, but neglecting the effect of the snow line in the inner disk.}
   \label{main_SEDs}
\end{figure}

\subsection{Spectral energy distributions}

Figure~\ref{main_SEDs} shows the resulting SEDs obtained from the dust density distributions illustrated in Fig~\ref{dust_distribution}. We assumed a disk inclination of 15$^\circ$ at a distance of 140~pc. The partial filtration and the change in $v_{\rm{frag}}$ at  $r_{\rm{snow}}$ have a clear effect on the shape of the SEDs. At 1Myr of evolution, independent of the mass of the embedded planet, there is still a significant amount of dust in the inner disk that contributes to the NIR emission, and therefore all cases exhibit a NIR excess, typical of pre-transition disks. This is the traffic jam effect:  It slows dust down enough to make a visible appearance in the innermost parts of the disk. After 5~Myr of evolution, in the case of a $1~M_{\rm{Jup}}$ mass planet, the NIR excess still  remains owing to the constant replenishment of dust from the outer to the inner disk, and the SED morphology is similar to the status at 1~Myr.  This is the tunnel effect:  dust travels through the gap as small grains,  which grow quickly and soon after drift inwards, remaining undetected.  The small dust that makes it through the gap has a small surface density and is optically thin.  It reappears at the end of the tunnel, because the changed mechanical properties of dust aggregates replenish small particles and slow down the radial drift motion, which is the traffic jam at the end of the tunnel.  In the 5 and 15$~M_{\rm{Jup}}$ planets this never happens because the flux of dust into the gap and inner disk is completely shut off at the trap. To identify the role of the snow line on the SED, Fig.~\ref{main_SEDs} also shows  SEDs obtained by assuming a $1~M_{\rm{Jup}}$ planet after 5~Myr of evolution and following the same procedure as before, but neglecting the effect of the snow line in the inner disk. In this case, there is no NIR excess.

\subsection{Snow line located further out}

\begin{figure*}
 \centering
   	\includegraphics[width=16cm]{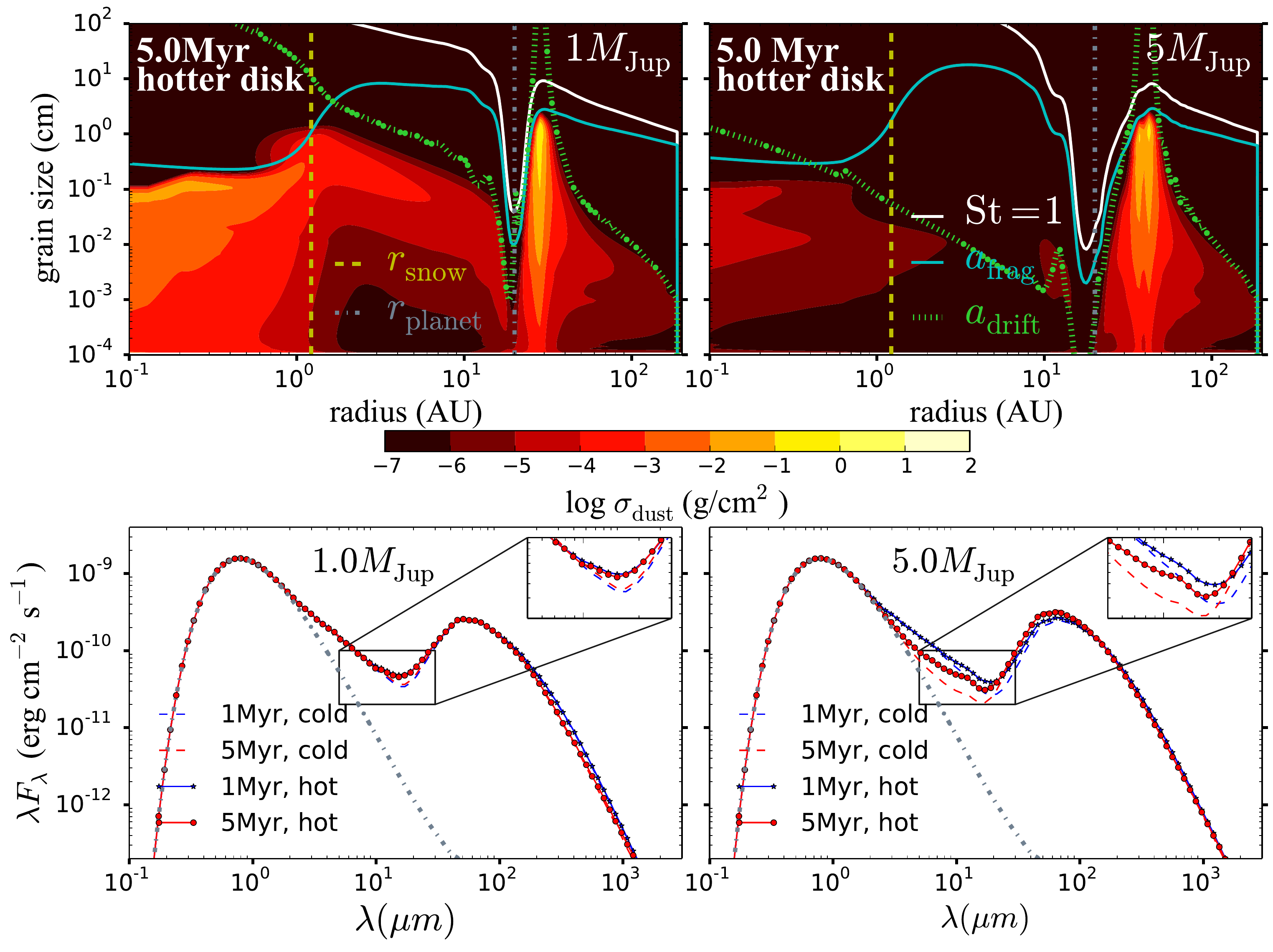}
   \caption{Vertically integrated dust density distribution after 5 Myr of evolution and the corresponding SEDs (after 1 and 5 Myr), considering a disk with higher temperature (hence $r_{\rm{snow}}$ =1.2~AU), for a planet mass of 1~$M_{\rm{Jup}}$ (left) and 5~$M_{\rm{Jup}}$ (right). For comparison, we plot the SEDs of the reference models (referred to as ``cold'').}
   \label{hotter_dust_SEDs}
\end{figure*}

To explore how the results are affected by moving the snow line further out, we increased the dust temperature profile such that $T_{\rm{snow}}{\simeq}200$K is at 1.2~AU, in agreement with the resulting mid-plane temperature from the radiative transfer simulations. Figure~\ref{hotter_dust_SEDs} shows the vertically integrated dust density distribution after 5~Myr of evolution and the corresponding SEDs when $r_{\rm{snow}}{\sim}1.2$~AU (``hot disk'') for 1 and 5$~M_{\rm{Jup}}$ mass planets. Because the variation of $v_{\rm{frag}}$ from 1 to 10 m~s$^{-1}$ occurs further out, the fragmentation barrier is below the drift barrier at ${\sim}$1.5~AU. As a result, the small grains produced  in the inner disk are distributed in a broader radial region. As in the case of the snow line at 0.4~AU, the inner part of the disk is practically empty of dust for the case of a 5$~M_{\rm{Jup}}$ mass planet. This broader dust distribution in the inner part does not have a significant effect on the NIR emission, but slightly affects the mid-IR (5-20~$\mu$m), in particular for the case of a 5$~M_{\rm{Jup}}$ mass planet at 5~Myr (Fig.~\ref{hotter_dust_SEDs}).

\section{Discussion}     \label{discussion}

In the context of planet-disk interaction combined with dust evolution and including the effect of the water snow line on the dust dynamics and growth, we showed that the NIR excess that characterises a subset of transition disks known as pre-transition disks is not necessarily an evolutionary effect. When the embedded planet causes partial filtration of particles (e.g. $1~M_{\rm{Jup}}$), there is a constant replenishment of small particles from the outer to the inner disk. The small particles in the outer disk are continuously reproduced at the pressure bump owing to fragmentation by turbulent velocities. When these small particles pass through the gap, they grow efficiently and are influenced by radial drift. However, as a result of the changes in the velocities of destructive collision in the absence of ice (inside the snow line), small particles are produced again. They are closely coupled to the gas and remain in the inner disk on long evolutionary time scales. As a consequence, in the case of a $1~M_{\rm{Jup}}$ mass planet the NIR excess remains for up to 5~Myr of evolution and the SED morphology remains almost identical (Fig.~\ref{main_SEDs}). Nonetheless, in the case of total filtration ($\gtrsim5~M_{\rm{Jup}}$), the inner part of the disk evolves from a filled to a practically empty dust cavity, completely removing any detectable NIR excess.

Our results are sensitive to the level of turbulence. On the one hand, with the same pressure gradient but higher disk turbulence, it is more difficult to retain particles inside pressure traps because of the stronger particle diffusion. On the other hand, increasing the disk turbulence decreases the pressure gradient. \cite{pinilla2012} showed for instance that for a  $1~M_{\rm{Jup}}$ mass planet and a turbulence of $\alpha_{\rm{turb}}=10^{-2}$, trapping does not occur at the outer edge of the planetary gap, hence a ring-shaped emission is not expected at mm wavelengths and the SEDs do not show a morphology typical of transition disks \citep[see also][Figure 6]{pinilla2015}. Thus, for a more turbulent disk, a more massive planet is required to enable particle trapping of the mm-sized grains. 

The grain size distribution in equilibrium depends on the size distribution of fragments after collision \citep{birnstiel2011}, which is usually assumed to be a power law, such that $n(m)dm\propto m^{-\xi}dm$. Numerical and laboratory experiments have shown that the typical values for $\xi$ are between 1 and 2. In this work we have assumed the value of 1.5. One may think that the filtration effect can be affected by the value of $\xi$. As an experiment, we lowered the value of $\xi$ to one for the case of a $1~M_{\rm{Jup}}$ mass planet. This implies fewer small particles when fragmentation occurs. The results do not significantly differ because small particles grow on very short time scales and overall dust distributions are similar after a million years of evolution. We presume that the reason why we do not reproduce the 10 $\mu$m feature is that at any radial location the dust density distribution of the small grains is much lower than the larger grains. 

As another experiment we investigated how sharper variations of $v_{\rm{frag}}$ affect the dust density distributions and the SEDs. We assumed that variations of $v_{\rm{frag}}$ from 1 to 10 m~s$^{-1}$ occur in the range between 0.3-0.6~AU  (versus a ${\sim}$2~AU transition in the presented models). In this case the fragmentation barrier is below the drift barrier in a smaller location (only inward of ${\sim}$0.5~AU), narrowing the distribution of small particles in the inner disk. This change has a marginal effect on the mid-IR emission, which decreases slightly.

One can think of other possible scenarios for the pre-transition disk geometry of a large outer disk combined with a tiny inner disk. One scenario could be that the inner disk region has completed its planetesimal formation process and contains hardly any dust. All the solids are in the form of planetesimals and planets. However, as we know from debris disks, occasional collisions can produce debris consisting of a collisional cascade of ever smaller bodies. Once the bodies become as small as metres, they drift rapidly towards the star. When they pass through the evaporation radius, they evaporate. Some of this vapour may be turbulently mixed outward, back behind the evaporation radius where it may condense out into micron-sized grains. This might cause a puff of smoke right at the evaporation radius, which would be detected as NIR flux. Models are needed to test if enough material is produced in the puff to reproduce the NIR excess. 

\section{Conclusions}     \label{conclusion}

We have studied the effect of changing the mechanical properties of dust aggregates at the water ice snow line in transition disks in which a gap has been carved out by a planet. Our findings are as follows:
\begin{itemize}
\item When icy dust aggregates pass through the snow line, their mechanical properties change. This might be due to the loss of ice in existing aggregates or to a complete disintegration of the aggregates followed by re-aggregation as ice-less aggregates. In either case, the maximum size of particles that is still stable against fragmentation in collisions strongly decreases, which causes all particles to be coupled to the gas more efficiently. The strong coupling dramatically decreases the inward drift motion of the particles, effectively leading to a traffic jam of dust between the snow line and the silicate sublimation point.

\item In disks where a gap has been carved by a very massive planet or companion (5~$M_{\rm{Jup}}$ or larger at 20~AU from the star), dust filtration at the planet-carved gap is very efficient and the inner disk is depleted of dust and gas. In this case, the traffic jam effect can maintain a detectable amount of dust inside the snow line for about one million years.

\item In disks where the gap has been carved by a lower-mass planet (only 1~$M_{\rm{Jup}}$), large grains remain trapped in the pressure bump, but small grains make it past the planet. These grains grow and drift quickly towards the star as relatively large grains and remain undetected on their way through the gap, as if they get to the snow line through a tunnel. Of course, inside the snow line the change of sticking properties leads to the traffic jam described above, making a warm component of dust visible. In the case of a low-mass planet, the pre-transition appearance can be maintained as long as the outer disk provides a source of dust and gas.

\item Radiative transfer calculations show that the SEDs of these models indeed resemble the overall appearance of pre-transition disks. Disks with massive planets lose the NIR excess after a few Myrs and look like true transition disks. Disks with less massive planets look like pre-transition disks for most of their remaining lifetimes.  

\end{itemize}

\begin{acknowledgements}
The authors are very grateful to E.~F.~van~Dishoeck, M.~Kama, and C.~Espaillat for all their enthusiasm and comments on this paper. We thank M.~Min and M.~de~Juan~Ovelar for all the feedback and help during the preparation of this manuscript. P.~P. is supported by Koninklijke Nederlandse Akademie van Wetenschappen (KNAW) professor prize to Ewine van Dishoeck.  T.~B. acknowledges support from NASA Origins of Solar Systems grant NNX12AJ04G. M.~B. acknowledges financial support from ``Programme National de Physique Stellaire" (PNPS) of CNRS/INSU, France.
\end{acknowledgements}

\bibliographystyle{aa}
\bibliography{ms.bbl}

\appendix

\section{Synthetic images and radial intensity profile} \label{appendix_a}

\begin{figure}
 \centering 
   \includegraphics[width=8.5cm]{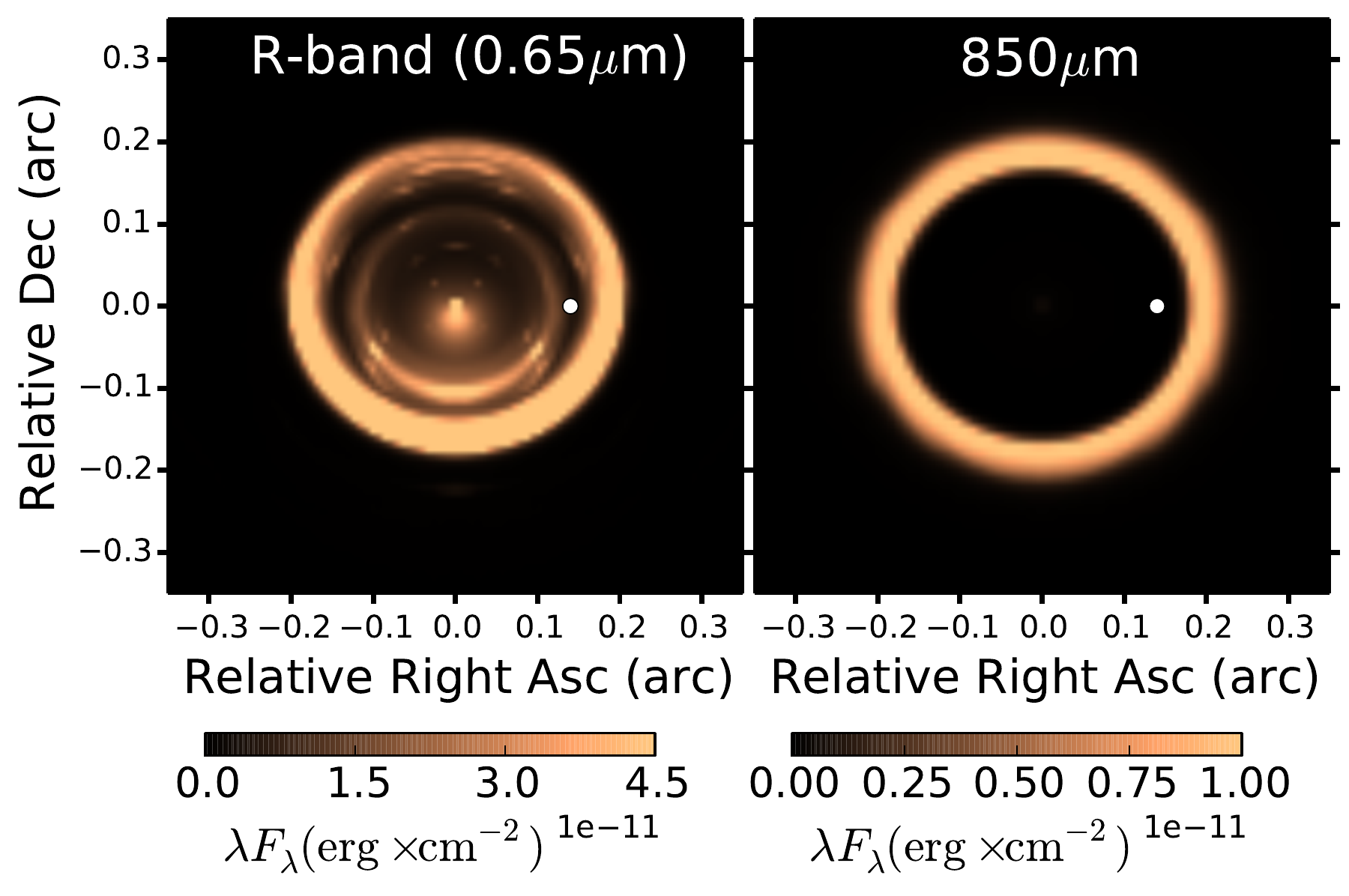}
    \includegraphics[width=8.5cm]{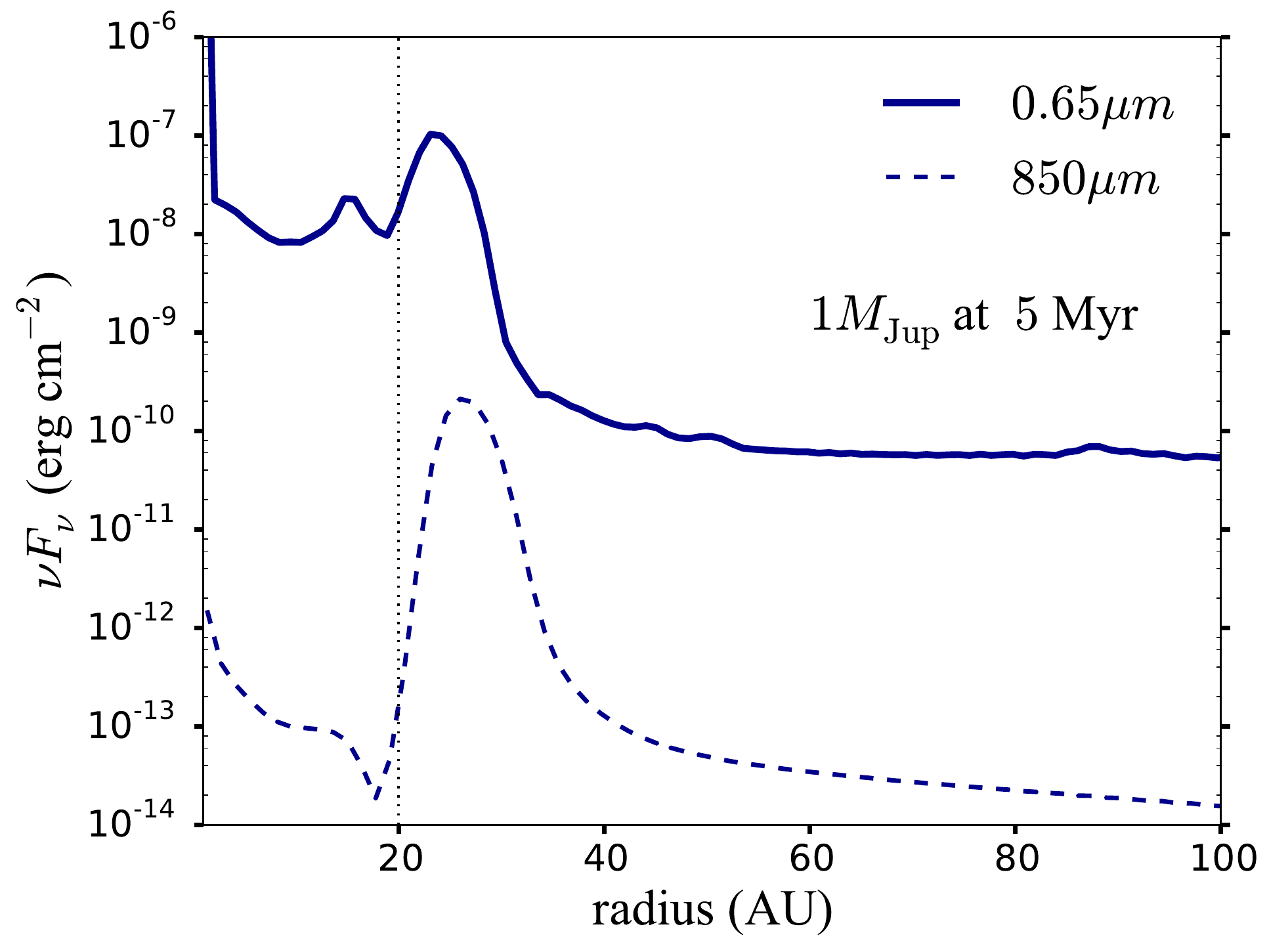}
   \caption{Simulated images (\emph{top}) and radial intensity profile (\emph{bottom}) at 0.65~$\mu$m and 850~$\mu$m, assuming a 1~$M_{\rm{Jup}}$ planet at 20~AU after 5~Myr of evolution.}
   \label{synthetic_images}
\end{figure}

Figure~\ref{synthetic_images} shows the simulated images and radial intensity profiles at 0.65~$\mu$m and 850~$\mu$m for the case of a 1~$M_{\rm{Jup}}$ mass planet embedded in the disk at 20~AU after 5~Myr of evolution, and considering the effect of the snow line on the dust evolution (Fig.~\ref{dust_distribution}). Large mm-sized grains only remain in the pressure trap at the outer edge of the gap, showing a single ring-like emission at mm wavelengths. The grains that go through the gap (\textit{tunnel}) remain invisible at mm wavelengths because they are small in size. The emission at R-band however is significant inside and outside the gap, as was previously shown by \cite{dejuanovelar2013}. Assuming a constant gas surface density,  \cite{gonzalez2015} show that if efficient growth happens at the location of the pressure trap (e.g. when fragmentation velocities are $\gtrsim30~\rm{m}~\rm{s}^{-1}$), a second  ring can be formed further out by considering the back-reaction of the dust on the gas. The recent ALMA image of the HL Tau disk \citep{alma2015} show multiple gaps and rings (${\sim} 7$), in which case multiple pressure bumps (which may or may not originate via planet-disk interaction) are needed.

\end{document}